\documentclass[aps,prl,showpacs,twocolumn,amsmath,amssymb,groupedaddress]{revtex4-1}
\usepackage{graphicx}
\usepackage{dcolumn}
\usepackage{bm}
\usepackage{color}
\usepackage{longtable}
\usepackage{multirow}

\begin{document}
\title{Crystal Structure, Lattice Vibrations, and Superconductivity of LaO$_{1-x}$F$_{x}$BiS$_{2}$}

\author{J. Lee$^{1}$}
\author{M. B. Stone$^{2}$}
\author{A. Huq$^{3}$}
\author{T. Yildrim$^{4}$}
\author{G. Ehlers$^{2}$}
\author{Y. Mizuguchi$^{5,6}$}
\author{O. Miura$^{5}$}
\author{Y. Takano$^{6}$}
\author{K. Deguchi$^{6}$}
\author{S. Demura$^{6}$}
\author{S.-H. Lee$^{1}$}

\affiliation{$^1$Department of Physics, University of Virginia, Charlottesville, VA 22904, USA}
\affiliation{$^2$Quantum Condensed Matter Division, Oak Ridge National Laboratory, Oak Ridge, Tennessee 37831-6393, USA}
\affiliation{$^3$Chemical and Engineering Materials Division, Oak Ridge National Laboratory, Oak Ridge, Tennessee 37831-6393, USA}
\affiliation{$^4$NIST Center for Neutron Research, National Institute of Standards and Technology, Gaithersburg, Maryland 20899, USA}
\affiliation{$^5$Department of Electrical and Electronic Engineering, Tokyo Metropolitan University, 1-1, Minami-osawa, Hachioji, 192-0397, Japan}
\affiliation{$^6$National Institute for Materials Science, 1-2-1, Sengen, Tsukuba, 305-0047, Japan}

\date{\today}

\begin{abstract}

Neutron scattering measurements have been performed on polycrystalline samples of the newly discovered layered superconductor LaO$_{0.5}$F$_{0.5}$BiS$_{2}$, and its nonsuperconducting parent compound LaOBiS$_{2}$. The crystal structures and vibrational modes have been examined. Upon F-doping, while the lattice contracts significantly along $c$ and expands slightly along $a$, the buckling of the BiS$_2$ plane remains almost the same. In the inelastic measurements, a large difference in the high energy phonon modes was observed upon F substitution. Alternatively, the low energy modes remain almost unchanged between non-superconducting and superconducting states either by F-doping or by cooling through the transition temperature. Using density functional perturbation theory we identify the phonon modes, and estimate the phonon density of states.  We compare these calculations to the current measurements and other theoretical studies of this new superconducting material. 
\end{abstract}

\maketitle

Superconductivity has been fascinating scientists for more than a century \cite{Ohnes}. Several families of superconducting (SC) materials and mechanisms have been found and proposed: notably type I or BCS superconductors where superconductivity is mediated by phonons, i.e. lattice vibrations, and type II superconductors whose superconductivity is yet to be fully understood. Many materials that belong to the latter category have layered crystal structures with low dimensionality. The cuprates with CuO$_{2}$ layers and the Fe-based superconductors with Fe-An (An: pnictogen or chalcogen anion) layers are two extensively studied examples \cite{Dai2012,Cuprate1,Cuprate2}. Very recently, a new family of materials based on BiS$_2$ layers has been found to be SC at low temperatures: Bi$_{4}$O$_{4}$(SO$_{4}$)$_{1-x}$ \cite{Bi4O4S3}, LaO$_{1-x}$F$_{x}$BiS$_{2}$ \cite{LaOBiS2}, NdO$_{1-x}$F$_{x}$BiS$_{2}$ \cite{NdOBiS2}, PrO$_{1-x}$F$_{x}$BiS$_2$ \cite{PrOBiS2}, and CeO$_{1-x}$F$_{x}$BiS$_2$ \cite{CeOBiS2}. The natural question that arises is whether the new Bi-based superconductors are type I BCS superconductors or another family of type II superconductors yielding a new route for unconventional superconductivity. 

Several theoretical studies have been reported especially for LaO$_{0.5}$F$_{0.5}$BiS$_{2}$ with the highest $T_c \approx 10.8$ K among the Bi-superconductors until now. Theoretical consensus so far is that in LaO$_{0.5}$F$_{0.5}$BiS$_{2}$ the Fermi level crosses conduction bands yielding electron pockets \cite{Minimal, Band, Theory, Phonon, Pairing}. From the quasi-one-dimensional (1d) nature of the conduction bands, these Fermi surfaces(FS) are nested by a ($\pi$,$\pi$) wave vector. The many body interactions and their relation to the superconductivity are, however, still controversial. A group of theories propose that FS nesting, just as in the Fe-pnictides, makes electronic correlations a candidate for the pairing mechanism \cite{Pairing,Impurity}. Alternatively, electron-phonon (e-ph) coupling constant ($\lambda$) calculations \cite{Theory,Phonon,Taner} find $\lambda \approx 0.85$ with $T_\text{c} \approx 11.3$~K, close to the experimentally found value. This suggests LaO$_{1-x}$F$_{x}$BiS$_{2}$ can be a conventional superconductor with strong e-ph coupling. Moreover, theories predict structural instabilities \cite{Theory,Taner} which places this system in close proximity to competing ferroelectric and charge density wave (CDW) phases \cite{Taner}. This instability is believed to be due to the anharmonic potential of the S ions lying on the same plane as the Bi ions. Just like the spin-density wave (SDW) instability in Fe-based superconductors, a CDW instability from negative phonon modes at or around the $M$ point, ($\pi$,$\pi$), is suggested to be essential to the superconductivity. The buckling of the BiS$_{2}$ plane, being related to the electronic structure and FS, was also suggested to be crucial in inducing superconductivity, and expected to decrease in the SC compound \cite{Taner}.

Experimentally, doping dependent structural measurements \cite{HP} suggest that the $a$ lattice parameter is relevant to superconductivity. Both $a$ and $T_c$ increase with F substitution, and become maximal near optimal doping $x\approx 0.5$. The existence of electron pockets near the FS was experimentally shown by Hall effect measurements \cite{Hall,Hall2}. Other experimental studies to test the theoretical scenarios, however, are limited. Thus, detailed experimental studies of the crystal structure and lattice vibrations of non-SC and SC compounds are crucial in understanding the new Bi-superconductors.

Here we report neutron diffraction and inelastic neutron scattering measurements both on non-SC LaOBiS$_{2}$ and SC LaO$_{0.5}$F$_{0.5}$BiS$_{2}$ to investigate how the crystal structure and phonon density of states (PDOS) change as a function of F-doping and temperature. We also performed density functional linear response calculations to identify the corresponding phonon modes. Experimentally obtained structural parameters are compared between the non-SC and SC compound and with calculated values. While significant differences in the phonon spectrum was observed at higher energy transfers upon F-substitution, little change is observed in the low energy portion of the spectrum. The calculations suggest significant modification in the low energy spectrum which should be relevant to the superconductivity, contrary to the experiment. We do not observe any meaningful change as a function of temperature in the phonon spectrum of the SC compound in the vicinity of $T_c$.

\begin{figure}[tp]
\includegraphics[width=0.90\hsize]{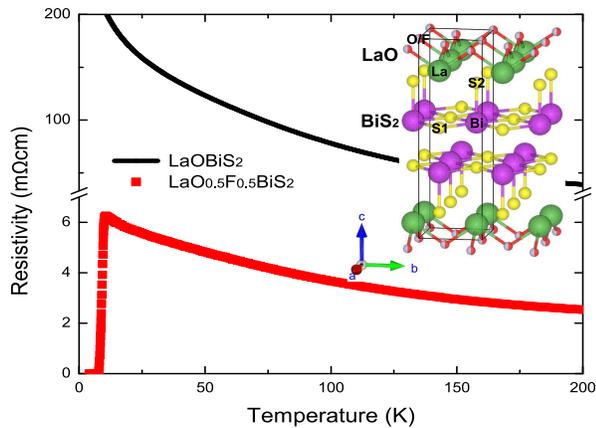}
\centering
\caption{(color online).\label{Fig01} Temperature dependent resistivity of LaOBiS$_{2}$ (black lines) and LaO$_{0.5}$F$_{0.5}$BiS$_{2}$ (red squares). The inset shows the crystal structure of LaO$_{1-x}$F$_{x}$BiS$_{2}$.}
\end{figure}

A 4.89 g polycrystalline sample of LaOBiS$_{2}$ was synthesized using the solid state reaction method under ambient pressure while a 0.89 g polycrystalline sample of LaO$_{0.5}$F$_{0.5}$BiS$_{2}$ was synthesized under high pressure at the NIMS, Tsukuba, Japan. Details concerning sample synthesis are described in Ref.~\cite{LaOBiS2,HP}.  Figure~\ref{Fig01} shows the low-temperature resistivity of the parent and F substituted compound.  LaO$_{0.5}$F$_{0.5}$BiS$_{2}$ has a clear SC transition at $T_c \approx 10.8$~K.

Neutron scattering measurements were performed at the Spallation Neutron Source (SNS) using the POWGEN diffractometer, the wide angular range chopper spectrometer (ARCS), and the Cold Neutron Chopper Spectrometer (CNCS).  Samples were loaded into vanadium (at POWGEN and CNCS), or aluminum (at ARCS) cans with a He atmosphere and mounted to the closed cycle refrigerator (at POWGEN and ARCS) or liquid helium cryostat (at CNCS). Neutron diffraction data were collected using a wavelength band to cover a wide range of d spacing from 0.55 to 4.12~$\text{\AA}$ at POWGEN \cite{POWGEN}.  Inelastic neutron scattering (INS) measurements were performed at ARCS \cite{ARCS} with monochromatic neutrons of incident energies E$_\text{i} = 40$~and 80~meV. For improved resolution at low energy transfer, INS measurements were performed at CNCS \cite{CNCS} with E$_{i}$ = 25 meV.  All the INS data presented here are corrected for background by subtraction of an empty can measurement.

The phonon calculations were performed using QUANTUM-ESPRESSO \cite{QE}. A $9\times 9 \times 3$ regular grid over the irreducible Brillouin Zone was used for the self-consistent calculation of the F substituted and parent compound. A manual check of convergence for grid density, energy cutoff, and lattice parameter values was performed before optimization of structural and atomic position values. To simulate the half doping in the SC sample, we replace oxygen at one of the $2a$ Wyckoff sites with F in an ordered fashion.

\begin{figure}[tp]
\includegraphics[width=0.90\hsize]{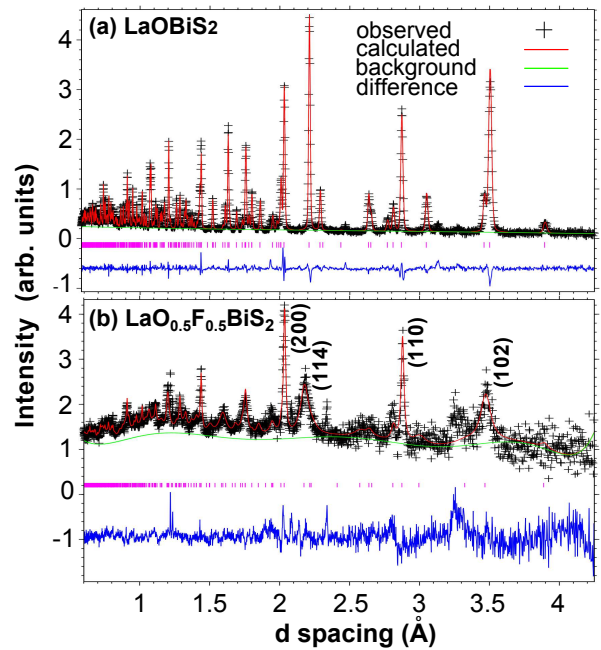}
\centering
\caption{(color online).\label{Fig02} Neutron powder diffraction data obtained from (a) LaOBiS$_{2}$, and (b) LaO$_{0.5}$F$_{0.5}$BiS$_{2}$ at 15K. Black crosses represent observed data.  Red, green, and blue solid lines are the calculated intensity, estimated background, and difference between the observed and calculated intensity obtained by GSAS \cite{Gsas}, respectively.}
\end{figure}

Figure~\ref{Fig02}(a) shows neutron diffraction data and Rietveld refinements obtained from LaOBiS$_{2}$. The nuclear Bragg peaks are instrumental resolution limited indicating good crystallinity. The data for the parent compound are well reproduced by the $P4/nmm$ space group with structural parameters and goodness of fit listed in Table~\ref{Tab1}. We determine the crystal structure, as shown in the inset of Fig.~\ref{Fig01}, to be very similar to that found by the preliminary x-ray scattering measurements \cite{LaOBiS2, HP}. The Bi ions form a square lattice just as the Cu and Fe ions do in the cuprates and Fe-based superconductors, respectively. There are two distinct Wyckoff sites for the sulfur ions, S$_1$ and S$_2$. While S$_1$ ions reside nearly on the same plane as the Bi square lattice with some buckling, S$_2$ ions are just above or below the Bi ions.

\begin{table}[tp] 
\squeezetable
\centering
\begin{tabular}{c c| c| c}
\hline\hline
      & & Experiment                   & Calculation \\
\hline
\multicolumn{2}{c|}{LaOBiS2} & wRp = 0.0799 & \\
\multicolumn{2}{c|}{P4/nmm T=15K} & $\chi^{2}$ = 13.48 & \\
\hline
\multicolumn{2}{c|}{a ($\text{\AA}$)}	& 4.05735(5) &	4.03949 \\
\multicolumn{2}{c|}{c ($\text{\AA}$)}	& 13.8402(3) &	14.30361 \\
La & 2c (0.5, 0, z) & 0.09065(9) & 0.08527 \\
Bi & 2c (0, 0.5, z)	& 0.3688(1)	& 0.36652 \\
S$_1$ & 2c (0.5, 0, z)	& 0.3836(3) &	0.39493 \\
S$_2$ & 2c (0.5, 0, z)	& 0.8101(2) &	0.81120 \\
O	 & 2a (0, 0, 0)	& $-$ & $-$ \\
\multicolumn{2}{c|}{$\left| \text{z$_{Bi}$}-\text{z}_{S1}\right|/\text{z}_{Bi}$} & 4.01(8) (\%) &	4.44373 (\%) \\
\hline\hline

\multicolumn{2}{c|}{LaO$_{0.5}$F$_{0.5}$BiS$_2$} & wRp = 0.0647 & \\
\multicolumn{2}{c|}{P4/nmm T=15K} & $\chi^{2}$ = 2.252 & \\
\hline
\multicolumn{2}{c|}{a ($\text{\AA}$)}	& 4.0651(3) &	4.07989 \\
\multicolumn{2}{c|}{c ($\text{\AA}$)}	& 13.293(7) &	13.42520 \\
La &	2c (0.5, 0, z) &	0.1007(5) & 0.10340 \\
Bi &	2c (0, 0.5, z) &	0.3793(5) & 0.38386 \\
S$_1$ &	2c (0.5, 0, z) &	0.362(2) & 0.38401 \\
S$_2$ &	2c (0.5, 0, z) &	0.815(1) &	0.81600 \\
O	& 2a (0, 0, 0) & $-$ & $-$ \\
F	& 2a (0, 0, 0) & $-$ & $-$ \\
\multicolumn{2}{c|}{$\left| \text{z$_{Bi}$}-\text{z}_{S1}\right|/\text{z}_{Bi}$} & 4.5(5) (\%) &	0.02413 (\%) \\
\hline\hline
\end{tabular}
\caption{\label{Tab1} Refined structural parameters of LaOBiS$_2$ and LaO$_{0.5}$F$_{0.5}$BiS$_{2}$ obtained from neutron powder diffraction and calculated with structural optimization implemented in QUANTUM ESPRESSO.  The quantity $\left| \text{z$_{Bi}$}-\text{z}_{S1}\right|/\text{z}_{Bi}$ characterizes the amount of buckling. wRp and $\chi^{2}$ respectively are the weighted R factor and chi squared values from the structural refinements \cite{Gsas}.}
\end{table}

On the other hand, for the SC LaO$_{0.5}$F$_{0.5}$BiS$_{2}$ most of the nuclear Bragg peaks are broader than instrumental resolution (Fig.~\ref{Fig02}(b)), indicating imperfect crystallinity. Similar broadening has been reported in prior X-ray measurements \cite{LaOBiS2,HP}. We note that the broad peaks have an asymmetric line-shape, characteristic of a low dimensional crystal ordering. The asymmetric broad peaks index with nonzero $l$ values, while $l=0$ Bragg peaks are considerably sharper and more symmetric. This is clearly seen in Fig.~\ref{Fig02}(b), for example, where the (200) peak at $d=2.03$~{\AA} and the (110) peak at $d=2.87$~{\AA} are sharp while the (114) peak at $d=2.17$~{\AA} and the (102) peak at $d=3.47$~{\AA} are broad and asymmetric. This suggests that strain may be induced along the $c$-axis due to a random replacement of the F ions at O sites. As a result, the La(O,F) planes are not well ordered along the $c$-axis. The solid line in Fig.~\ref{Fig02}(b) is our best refinement to the data, where a phenomenological model of anisotropic broadening has been used \cite{Broadening}. This coarsely reproduces the diffraction data and allows for reasonable determination of the structural parameters, summarized in Table ~\ref{Tab1}.

\begin{figure}[tp]
\includegraphics[width=0.90\hsize]{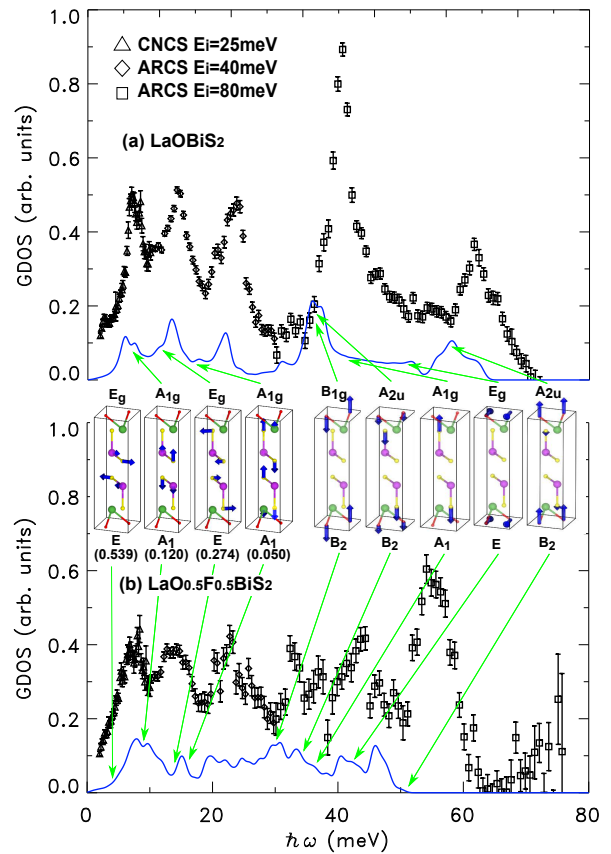}
\centering
\caption{(color online).\label{Fig03} GDOS as a function of $\hbar\omega$ for the (a) parent non-SC and (b) F substituted SC compound. Black symbols are the experimental data; blue solid lines are the calculated GDOS scaled for comparison. Three measurements have been combined: below 10 meV are from CNCS E$_{i}$ = 25 meV, between 10 and 30 meV from ARCS E$_{i}$ = 40 meV, and above 30 meV from ARCS E$_{i}$ = 80 meV. To obtain the GDOS, the neutron scattering intensity was averaged over momentum transfers of [3, 6] $\text{\AA}^{-1}$, [4, 7] $\text{\AA}^{-1}$, and [3, 8.5] $\text{\AA}^{-1}$, respectively. The phonon modes at the $\Gamma$ point are shown in the inset with their symmetry for the parent (above) and SC (below) compound. Atoms are shown with the same colors as in Fig.~\ref{Fig01}, and their relative displacements are represented with thick blue arrows. The left four figures in the inset are the vibrational modes that are theoretically expected to have the large e-ph couplings, with each $\lambda$ shown in parentheses. For the first mode shown here, the corresponding energy becomes negative ($-2.9$~meV) in the parent compound. For doublet modes, only one of the two orthogonal modes is shown.}
\end{figure}

We find upon substitution with F, the lattice elongates along $a$ by $\approx 0.2 \%$ while it contracts along $c$ by $\approx 4.1 \%$, consistent with previous x-ray results \cite{HP}. This F-doping induced lattice change is reproduced by our calculations although the rates are different. The buckling of the BiS$_2$ plane is found to slightly increase as shown in Table~\ref{Tab1}.  This contradicts a prior theoretical prediction \cite{Taner} as well as our own calculations.

We note that there is discrepancy between experimental and calculated structural parameters.  The discrepancy in the $z$ position of the S$_1$ atom was regarded as a sign of a possible structural instability in this system \cite{Taner}. This deviation becomes larger with F substitution. According to frozen phonon calculations, the instability at or near the $M$ point can reduce the symmetry of the F substituted compound statically to P222$_1$ \cite{Taner}, the CDW phase. This structural transition, however, could not be experimentally confirmed due to the Bragg peak broadening. 

To examine if the pairing mechanism of the superconductivity is phononic, we have performed INS measurements to obtain the neutron weighted generalized PDOS (GDOS). Figure~\ref{Fig03} shows the GDOS as a function of energy transfer, $\hbar\omega$, for the two compounds at $T=5$~K. For the non-SC LaOBiS$_{2}$, (Fig.~\ref{Fig03}(a)), there are well-defined phonon modes over a wide range of $\hbar\omega$ up to 70 meV. At least five prominent bands of lattice vibrations are present at 7.7(2), 14.2(1), 22.9(2), 40.4(1), and 61.8(4) meV. Upon F-substitution, as shown in Fig.~\ref{Fig03} (b), the phonon modes at higher energies change significantly. All the vibrational modes become broader than their corresponding modes of LaOBiS$_{2}$. Furthermore, the top of the band softens in energy from 61.8(4) to 55.2(1) meV, and the sharp 40 meV peak significantly weakens and broadens. Conversely, the first two low energy peaks remain similar to those of the parent compound even though they broaden.

\begin{figure}[tp]
\includegraphics[width=0.90\hsize]{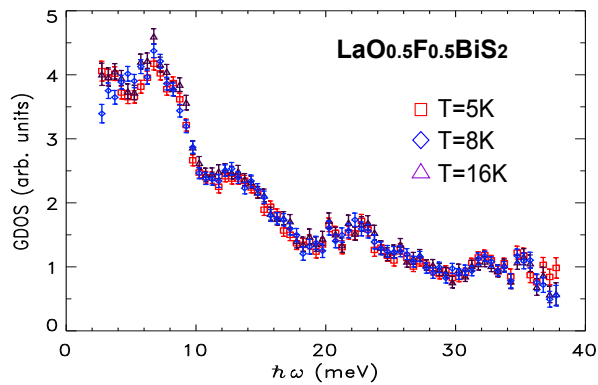}
\centering
\caption{(color online).\label{Fig04} Temperature dependence of the dynamical susceptibility. The data have been measured at ARCS with E$_{i}$ = 40 meV, and integrated over Q = [4, 7] $\text{\AA}^{-1}$.}
\end{figure}

Our calculated GDOS are shown as solid lines in Fig.~\ref{Fig03}. For the non-SC sample, it reproduces the observed prominent phonon modes reasonably well. Our calculations at the zone center show that the 40 and 62 meV bands are mainly due to vibrations of light ions. For example, in the insets of Fig.~\ref{Fig03}, the last 5 figures show the highest energy vibration modes.  These modes mostly involve O and/or S$_2$ vibrations. The intermediate energy modes around 13 and 23 meV are mainly due to S$_1$ and/or S$_2$ vibrations, while the low energy modes below $\approx 10$~meV are due to the vibrations of the BiS$_{2}$ and/or LaO layers. Our calculations at the zone center also yield imaginary unstable phonon modes involving vibrations of the BiS$_2$ layer, which was previously reported to be the sign of anharmonic ferroelectric soft phonons \cite{Taner}.

For the SC-sample, the calculated GDOS reproduces the features occurring at large $\hbar\omega$ (albeit shifted in energy); the broadening of high energy peaks, reduction of the sharp 40 meV peak, and softening of the highest energy O/F vibration modes. These changes can be partially understood as being due to the high energy O modes being shifted because of substitution with heavier F ions. This, however, fails to explain the low energy data. According to the calculation, most of the e-ph coupling comes from the low energy modes below $\approx 20$~meV, and as such this is where we expect meaningful changes of the GDOS relevant to the superconductivity to occur. While the theory predicts considerable re-distribution of GDOS in this low energy region, we do not observe any such change in the measured spectrum upon F-substitution.

To further probe for a phonon anomaly associated with superconductivity, we also examine the temperature dependence of the dynamical susceptibility, $\chi^{''}(\omega)$, of the SC LaO$_0.5$F$_0.5$BiS$_2$. Figure~\ref{Fig04} shows $\chi^{''}(\omega)$ at three different temperatures spanning $T_\text{c} = 10.8$~K. Our experimental data show that $\chi^{''}(\omega)$ does not change within the experimental errors when the system transits from the normal to SC state.

In summary, our crystal structure refinements show that upon F-doping the buckling of the BiS$_2$ plane remains very similar to the parent compound. This is inconsistent with a theoretical study that the buckling decreases upon F-doping, the change of which facilitates an electronic environment for the appearance of the superconductivity \cite{Taner}. Furthermore, it was theoretically predicted that, in the SC phase, a significant change in the phonon density of states at low energies would occur due to a possible large e-ph coupling. Our inelastic neutron scattering data, however, yield no considerable change in the low energy phonon modes as the system becomes SC either by F-doping or by cooling through the superconducting transition. Our results should provide important constraints on future theoretical works examining these new Bi-based superconductors.

This research at UVA and ORNL's Spallation Neutron Source were sponsored by the Division of Materials Sciences and Engineering, Basic Energy Sciences (BES), US Department of Energy (DOE) under Award No. DE-FG02-10ER46384, and by the Scientific User Facilities Division, BES, US DOE, respectively.



\end{document}